\documentclass[a4paper]{jpconf}
\usepackage{graphicx,threeparttable}
\begin{document}
\title{Energy spectra of $^3$He-rich solar energetic particles associated with coronal waves}

\author{R Bu\v{c}\'ik$^{1, 2}$, D E Innes$^{2, 3}$, G M Mason$^4$ and M E Wiedenbeck$^5$}
\address{$^1$ Institut f\"{u}r Astrophysik, Georg-August-Universit\"{a}t G\"{o}ttingen, D-37077, G\"{o}ttingen, Germany}
\address{$^2$ Max-Planck-Institut f\"{u}r Sonnensystemforschung, D-37077, G\"{o}ttingen, Germany}\address{$^3$ Max Planck/Princeton Center for Plasma Physics, Princeton, NJ 08540, USA}
\address{$^4$ Applied Physics Laboratory, Johns Hopkins University, Laurel, MD 20723, USA}
\address{$^5$ Jet Propulsion Laboratory, California Institute of Technology, Pasadena, CA 91109, USA}

\ead{bucik@mps.mpg.de}

\begin{abstract}
In addition to their anomalous abundances, $^3$He-rich solar energetic particles (SEPs) show puzzling energy spectral shapes varying from rounded forms to power laws where the later are characteristics of shock acceleration. Solar sources of these particles have been often associated with jets and narrow CMEs, which are the signatures of magnetic reconnection involving open field. Recent reports on new associations with large-scale EUV waves bring new insights on acceleration and transport of $^3$He-rich SEPs in the corona. We examined energy spectra for 32 $^3$He-rich SEP events observed by ACE at L1 near solar minimum in 2007-2010 and compared the spectral shapes with solar flare signatures obtained from STEREO EUV images. We found the events with jets or brightenings tend to be associated with rounded spectra and the events with coronal waves with power laws. This suggests that coronal waves may be related to the unknown second stage mechanism commonly used to interpret spectral forms of $^3$He-rich SEPs. 
\end{abstract}

\section{Introduction}
Small $^3$He-rich solar energetic particle (SEP) events remain $\sim$40 years after their discovery one of the less explored phenomenon in solar physics. The anomalous abundances of $^3$He-rich SEPs have been explained by various mechanisms including gyro-resonant interactions with plasma wave turbulence \cite{fis78, mil98} or interactions with multiple magnetic islands \cite{dra12}.

The energy spectra of major ion species in some $^3$He-rich SEP events are power laws or broken power laws but in other events $^3$He and Fe exhibit rounded spectra toward the low energies with $^3$He rollovers at $\sim$100-600\,keV\,nucleon$^{-1}$ and Fe rollovers below $\sim$100\,keV\,nucleon$^{-1}$ \cite{mas00, mas02}. It has been suggested that rounded spectra arise from the basic mechanism of $^3$He enrichment and power laws involve a further stage of acceleration. It is well known that power law spectra are produced by shock acceleration models (see e.g. \cite{ver15}, for a review) but shock detection is not typically reported in $^3$He-rich SEP events (e.g., via type II radio bursts). Notice that the spectral shape can deviate from power law when shock propagates into an inhomogeneous solar wind \cite{zan00}.

$^3$He-rich SEP sources have been associated with EUV jets and jet-like CMEs \cite{kah01, wan06, nit06}. These are signatures of magnetic reconnection involving open field \cite{shi92} which may create turbulence and magnetic islands for a particle acceleration. With new observing capabilities on STEREO and SDO it has been recently shown that some $^3$He-rich SEP events were associated with large-scale coronal EUV waves \cite{nit15, buc15}. These waves were found without CMEs or only with weak coronal outflows. A common view on EUV waves is that they present or contain magnetosonic waves which can even steepen to the shocks (see \cite{war15}, for a review). 

The aim of this paper is to examine relation between energy spectral characteristics of $^3$He-rich SEPs and their flare signatures in the corona. 

\section{Observations and discussion}
\begin{figure}
\begin{center}
\includegraphics[width=\textwidth]{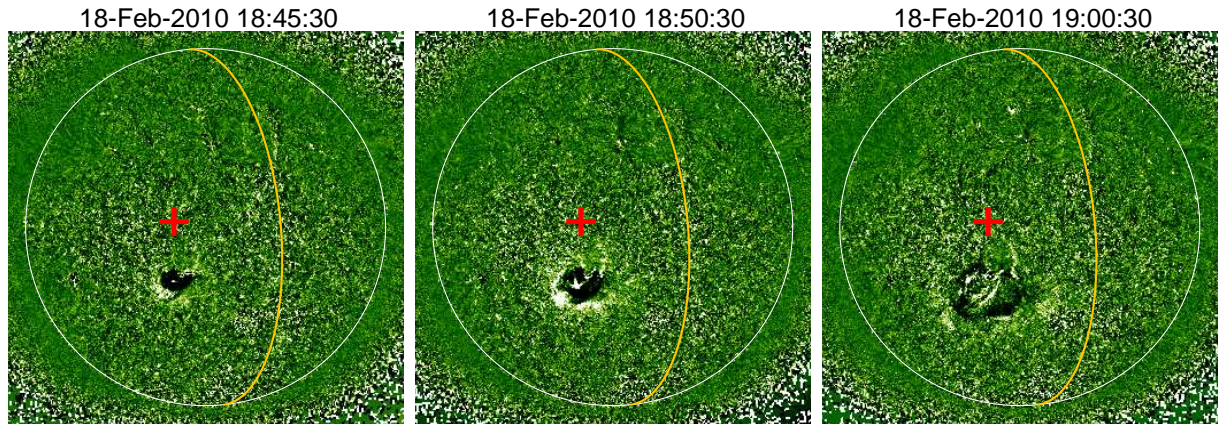}
\end{center}
\caption{\label{fig1}STEREO-A 195 {\AA} EUV 2.5 min running differences images of the solar disk on 2010 February 18. Yellow arcs indicate the west limb from the Earth view and red pluses the L1 magnetic foot-point. At this time STEREO-A was separated from the Sun-Earth line by 65$^{\circ}$.}
\end{figure}

\begin{figure}
\begin{center}
\includegraphics[width=\textwidth]{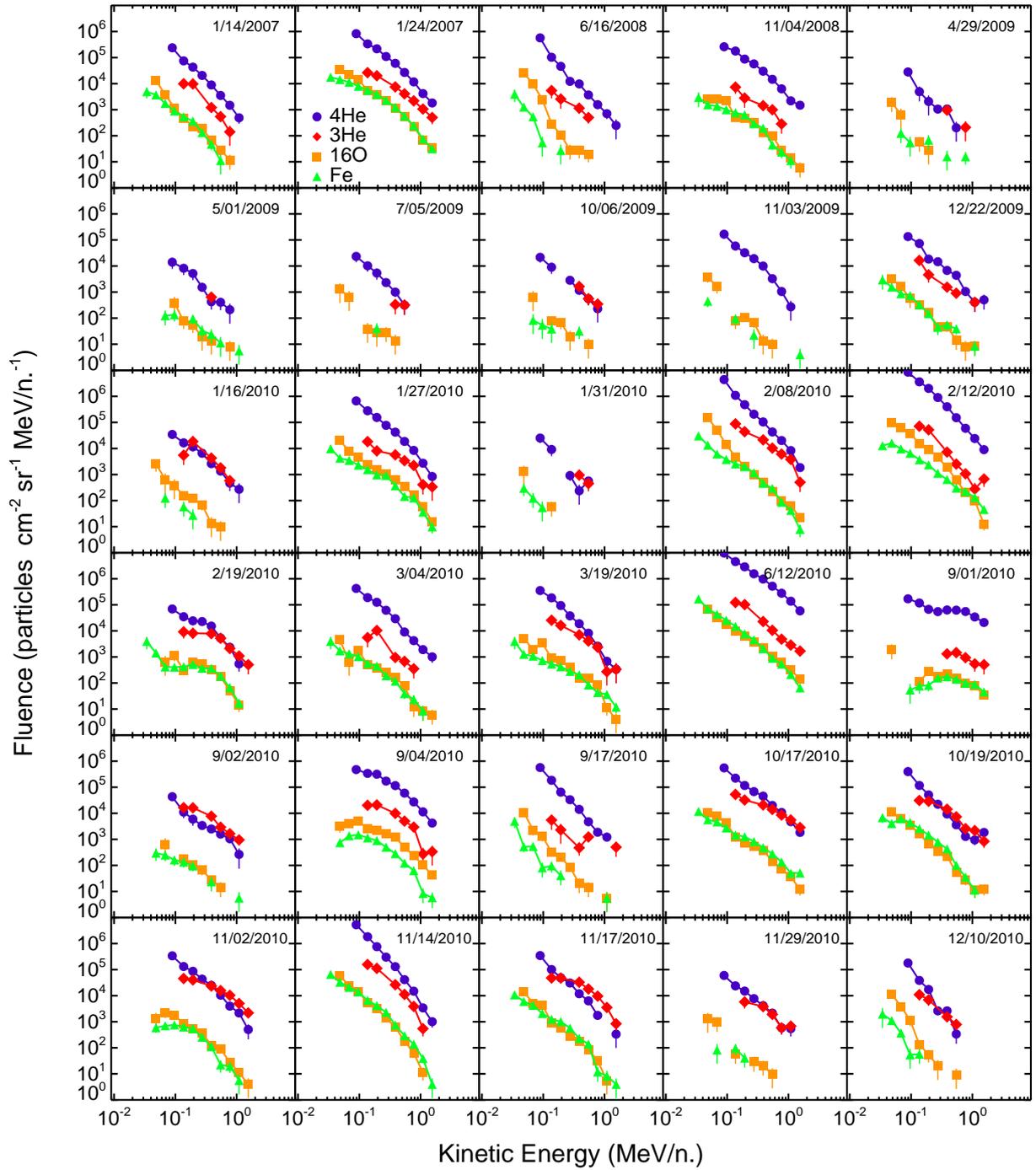}
\end{center}
\caption{\label{fig2}Energy spectra of $^4$He, $^3$He, O, Fe at $\sim$0.03-2\,MeV\,nucleon$^{-1}$ observed with ACE/ULEIS for 30 $^3$He-rich SEP events in 2007-2010.}
\end{figure}

We examine energy spectra in 32 $^3$He-rich SEP events observed at L1 during solar minimum conditions in 2007-2010. The events have been examined in detail in our recent survey \cite{buc16}. Four events contain a high energy ($>$25\,MeV) but relatively low intensity solar proton component \cite{ric14}, which might not be typical $^3$He events but rather large gradual events or a mix of both. Three of these events show still very high $^3$He enrichment and all have high Fe/O ratio, typical of $^3$He-rich SEPs. Note that such cases if not identified may present one source of variability in $^3$He-rich SEP events. Table~\ref{tab1} lists all the investigated events. Column 1 gives the event number, column 2 the event start day, column 3 the $^3$He energy spectral form at $\sim$0.1-2\,MeV\,nucleon$^{-1}$, column 4 indicates whether the ion event had velocity dispersion, column 5 the EUV flare shape and column 6 the CME speed reported in SOHO/LASCO catalog. The EUV flares were examined using STEREO/EUVI \cite{how08}; the energetic ions were detected by ACE/ULEIS \cite{mas98} and/or ACE/SIS \cite{sto98}. Figure~\ref{fig1} shows an example of EUV wave associated with $^3$He-rich SEP event on 2010 February 19. Shown are running differences images at three times to illustrate the wave expansion. Figure~\ref{fig2} presents the event integrated $^4$He, $^3$He, O, Fe fluence energy spectra for 30 events. The spectra for two remaining events (2007-May-23 and 2008-Feb-4) have been shown in \cite{mas09}. The energetic ion fluences for events in 2009 were very weak. Specifically the ULEIS did not observe statistically significant $^3$He fluences in 2009 November 3 event though at higher energies the SIS observed a $^3$He-rich SEP event \cite{wie13}. 

\begin{table}
\centering
\begin{threeparttable}
\caption{\label{tab1}$^3$He-rich SEP events.}
\begin{tabular}{clcccc}
\br
\#&Start day&$^3$He spectrum\tnote{a}&Dispersion&EUV flare\tnote{b}&CME\tnote{c}\\
\mr
1&2007-Jan-14&PL&no&B&...\\
2&2007-Jan-24&R&yes&B&...\\
3&2007-May-23&PL&no&W&544\\
4&2008-Feb-4&?&no&B&...\\
5&2008-Jun-16&?(PL)&?&J&...\\
6&2008-Nov-4&?(PL)&yes&W&732\\
7&2009-Apr-29&...&?&W&239\\
8&2009-May-1&...(R)&?&W&...\\
9&2009-Jul-5&...&?&W&yes\\
10&2009-Oct-6&PL&?&B&...\\
11&2009-Nov-3&...&?&W&yes\\
12&2009-Dec-22&PL&no&W&318\\
13&2010-Jan-16&PL&yes&J&...\\
14&2010-Jan-27&?(PL)&yes&W&...\\
15&2010-Jan-31&...(PL)&?&W&219\\
16&2010-Feb-8&PL&no&W&yes\\
17&2010-Feb-12&PL&no&W&509\\
18&2010-Feb-19&R&yes&W/B&223\\
19&2010-Mar-4&PL&yes&W&374\\
20&2010-Mar-19&PL&yes&W&...\\
21&2010-Jun-12&PL&no&W&486\\
22&2010-Sep-1&R&no&W&1304\\
23&2010-Sep-2&R&no&J&...\\
24&2010-Sep-4&R&yes&B&...\\
25&2010-Sep-17&?&no&J&yes\\
26&2010-Oct-17&PL&yes&W&304\\
27&2010-Oct-19&R&yes&B&385\\
28&2010-Nov-2&R&yes&W&253\\
29&2010-Nov-14&R&no&B&442\\
30&2010-Nov-17&R&yes&J&639\\
31&2010-Nov-29&R&?&W&505\\
32&2010-Dec-10&PL&?&W&299\\
\br
\end{tabular}
\begin{tablenotes}
\item[a] at 0.1-2\,MeV\,nucleon$^{-1}$; PL-power law, R-rounded, ?-unclear; the spectral form in parentheses is for Fe and extends down to 30\,keV\,nucleon$^{-1}$; spectra in events \#2-6, 14, 26-30 reported in \cite{wie00,mas09, buc15, nit15} 
\item[b] B-brighthening, J-jet, W-wave
\item[c] CME speed in km\,s$^{-1}$ from LASCO catalog (cdaw.gsfc.nasa.gov/CME\_list); yes-CME observed with STEREO/COR-1  
\end{tablenotes}
\end{threeparttable}
\end{table}

We are able to assign the $^3$He spectral shape for 22 from 32 investigated $^3$He-rich SEP events (see Table~\ref{tab1}). In the remaining 10 events $^3$He spectral points show complex variations (5 events) or we have insufficient number of spectral points (2 points as a maximum). The power law spectrum has been found in 12 events, where 3 events are with jet/brightening and 9 with a coronal wave. The rounded spectral shape is found in 10 events, where 6 are with jet/brightening and 4 with a wave. Since in the two class scheme the Fe spectrum behaves as $^3$He, we examined the Fe spectrum in the cases where $^3$He is unclear. This adds 5 more events. We found 16 events with $^3$He or Fe having power laws where 4 events were with jet/brightening and 12 with a coronal wave. This confirms the previous finding depending  solely on $^3$He spectrum. Including Fe we now have 11 events with rounded spectra where  6 are with jet/brightening and 5 with a wave, again consistent with the previous result. This shows that events with jet/brightening have a slight tendency to have rounded $^3$He spectra (6 of 9) and events with a wave to have power law $^3$He spectra (9 of 13). This is consistent with a recent study reporting similar tendency for 13 $^3$He-rich SEP events with jets and 4 events with EUV waves \cite{nit15}. 

Excluding 4 events with high energy solar proton components (\#12, 17, 21, 22), which are associated with EUV waves and all but one have power law $^3$He spectra, the results will not change dramatically. The $^3$He-rich SEP events with EUV waves have still a tendency to have power law $^3$He spectra (6 of 9) or power law $^3$He or Fe spectra (9 of 13). Exclusion of all events with a high energy proton component may be too conservative as for example one such event, 2009-Dec-22, have very high $^3$He and Fe enrichments (386 keV\,nucleon$^{-1}$ $^3$He/$^4$He$\sim$0.23, Fe/O$\sim$1.16). The 2 of 3 above events with the proton component which do not have an occulted source active region (AR) have reported type II radio bursts (ftp.ngdc.noaa.gov). The CMEs which accompany these solar proton events were generally not too fast (318, 486, 509, 544, and 1304 km\,s$^{-1}$ as reported by LASCO catalog). The slowest one in 2009-Dec-22 event was also associated with a type II radio burst. We get similar results if we consider coronal wave events only with slow CMEs ($<$350 km\,s$^{-1}$) or with no CME: $^3$He power law spectrum in 5 from 7 events and $^3$He (or Fe) power law in 7 from 10 events. 

In our earlier study \cite{buc15} we report rounded $^3$He, Fe spectra in the event with a weak EUV wave, probably without a significant influence on energetic ions. Therefore we look at the wave associated events with $^3$He rounded spectra (2010-Feb-19, 2010-Sep-1, 2010-Nov-2, 2010-Nov-29). The coronal wave in February 19 event was ejected about 30 min earlier, not at the time of the type III radio burst. Note this is the only case in this survey when the wave was not co-temporal with the flare and/or type III burst. Though overall the wave looks quite significant in the February 19 event (see Figure~\ref{fig1}), maybe it was ejected too early to have an effect on the later particle injection. The wave in the September 1 event had a large spatial scale but faded before reaching the L1 foot-point. The February 19 and September 1 events have all $^4$He, $^3$He, O, Fe spectra rounded, the feature has not been previously reported (see later). The November 2 and November 29 events are associated with quite narrow waves with less bright fronts. Furthermore the November 2 wave has the narrowest front in this study and the source AR in the November 29 event was extraordinary separated from the nominal magnetic connection. Thus the influence of the waves on energetic ion spectra in these two events would be minimal.

We have also 3 events with power law $^3$He spectra (2007-Jan-14, 2009-Oct-6, 2010-Jan-16) associated with a brightening or a jet. The October 6 event spectrum could be still rounded as we do not have spectral points below 400 keV\,nucleon$^{-1}$. Other two events have one, the lowest energy (100 keV\,nucleon$^{-1}$) spectral point, deviated from the power law dependence but this can be an instrumental issue. It has been discussed that propagation through interplanetary space is probably not a cause of a common spectral shapes observed in $^3$He-rich SEP events \cite{mas02}. Indeed, the 2007 January 14 event is a short duration event with a non-dispersive onset while the 2010 January 16 event is a dispersive one, though both have similar, power law, spectral shapes. The 2009 October 6 event is too weak to observe a velocity dispersion. In this study about half of the events with rounded $^3$He spectra have a dispersive onset (6 of 10; 1 event with unclear dispersion) indicating nearly scatter-free propagation and half of the events with power law $^3$He spectra have non-dispersive onsets (6 of 12; 2 events are unclear).

We found several $^3$He-rich SEP events which differ from the two class scheme introduced in \cite{mas00}. For example the following two events have only the $^3$He spectrum rounded, while the Fe is power law: the 2010 November 17 event and maybe also the 2010 November 14 event. Neither of these is associated with a coronal wave and both have the same solar source. Since we could not see this feature in other events (and no other reports are known) this may be due to specific conditions in a particular solar source region. However note on some other events where Fe shows a clear power law shape, $^3$He looks more complex (e.g., 2008-Nov-4, 2010-Jan-27) which might be of the previous type. Furthermore we found other events where all $^4$He, $^3$He, O, Fe spectra are rounded. Typical examples are the events of 2010 September 1 (already mentioned) and 2010 September 4. The September 1 event is particularly unusual, with the $^4$He spectrum continuing to rise at the lowest energies after the turnover. These two events are associated with different ARs: September 1 with a coronal wave and a high speed CME (with $>$25\,MeV proton event) and September 4 with a brightening. There is possibly another $^3$He-rich SEP event with all spectra rounded, the above mentioned 2010 February 19 event. Here after a turnover the Fe spectrum continues to rise at the lowest energies. This event was associated with a brightening but the wave ejected earlier was still present during the type III burst. Rounded $^4$He spectra have been predicted by a MHD turbulence model at lower ($<$20\,keV\,nucleon$^{-1}$) energies \cite{liu06} not covered by available instruments. 

\section{Summary}
We examined energy spectra for 32 $^3$He-rich SEP events and identified $^3$He spectral shapes for 22 events. We found a slight tendency for events with jets/brightenings in their solar source to have a rounded $^3$He spectrum and for events with coronal waves in their source to have a power law $^3$He spectrum. Though more events need to be examined to confirm this tendency the second stage mechanism behind the $^3$He-rich SEP acceleration may be related to coronal EUV waves even when CMEs are not seen or are represented only by slow white-light outflows. 

Half of the events with rounded $^3$He spectra (5 of 10) show peculiar behaviour not previously reported. We noticed events where clearly only the $^3$He spectrum is rounded and also events where all major species show rounded spectra. This may present further constraints on modelling of ion acceleration/propagation in $^3$He-rich SEP events.

\ack
The work of RB is supported by the Deutsche Forschungsgemeinschaft (DFG) under grant BU 3115/2-1. RB thanks Gary Zank for the invitation to the 15th AIAC that led to this paper.

\end{document}